\documentclass[a4paper]{article}
\usepackage[dvips]{graphicx}
\parskip=\baselineskip       % blank line between each paragraph.
\usepackage{url}
\usepackage{graphicx}
\usepackage{amsmath}

\begin{document}
\vspace*{-3cm}
\begin{flushright}
%CDF/STAT/PUBLIC/abcd\\
\today

\end{flushright}

%\vspace{0.5in}

\vspace{1.0ex}

\begin{center}
  
\begin{Large}
  {\bf A Paradox about the Distributions of \\ Likelihood Ratios?}
  
\end{Large}

\end{center}

%\vspace{0.5in}
\begin{center}

\vspace{5.0ex}

%\href{mailto:harper@fnal.gov}{Sam Harper}\footnote{The }\\ 
\vspace{1.0ex} {\Large Louis Lyons}\\
\vspace{1.0ex}
\emph{Blackett Lab., Imperial College, London SW7 2BW, UK} \\ 
 and\\
\emph{Particle and Astrophysics, Oxford OX1 3RH, UK}

\vspace{0.3cm}
e-mail: l.lyons@physics.ox.ac.uk
\vspace{1.0ex} 
\end{center}

\vspace{5.0ex}

\begin{center}
  
\begin{Large}
  {\bf  Abstract}
  
\end{Large}

\end{center}

%\today
%\end{flushright} 

We consider whether the asymptotic distributions for the log-likelihood ratio test statistic
are expected to be Gaussian or $\chi^2$. Two straightforward examples provide 
insight on the difference.

\section{Introduction}
\label{Introd}
In analysing data, the likelihood function provides a very useful method of 
determining free parameters, and also their uncertainties, 
in any model that is fitted to the data. The input to a likelihood calculation 
can be individual observations (unbinned likelihood), 
or histograms (binned likelihood); it is the former that we largely consider here.
Likelihoods are also used as part of 
the procedure to determine the Bayesian posterior probability.

Likelihood ratios (i.e. the ratio of the likelihood for two possible models) are 
used very much in Hypothesis Testing, where we compare how well the two
different hypotheses fit the data. In assessing their numerical values for a
particular data set,  is useful to know what are their 
expected asymptotic distributions for a specific hypothesis.

The Central Limit Theorem (CLT) can be used for this. It states that, 
if we take a linear combination of enough independent random variables $r_i$,
the distribution of this sum tends to a Gaussian. The $r_i$ can be from 
a single distribution, or from several several different distributions, 
provided there are enough samples from each distribution. 
The variances of each of these distributions must be finite.

The relevance of the CLT is as follows. To construct a likelihood function,
we first need the probability density function (pdf) $p(x|\mu)$ for obtaining data
$x_i$ for a given value of the theoretical parameter $\mu$.\footnote{Both $x$ and 
$\mu$ can be multi-dimensional.} Here $x_i$ denotes a series of independent and
identically distributed measurements of a single physical quantity $x$. Then
the likelihood ${\mathcal L}(\mu)$ is given by
\begin{equation}
{\mathcal L}(\mu) = \Pi\  p(x_i|\mu),
\end{equation}   
where the product extends over the different and independent $x_i$.
Its logarithm is
\begin{equation}
\ln {\mathcal L} = \Sigma \ln p(x_i|\mu)
\end{equation} 
It thus appears that the CLT applies to $\ln \mathcal L$, and similarly to the logarithm of a
likelihood ratio (LLR), and so their distributions should be Gaussian.

In contrast, a mathematical $\chi^2$ distribution is obtained for a variable 
\begin{equation}
S = \Sigma q_i^2
\end{equation}
where each of the $q_i$ is randomly and independently 
Gaussian distributed with zero mean and unit variance. If there are $K$ 
terms in the summation, $S$ will have a $\chi^2$ distribution with $K$ degrees
of freedom. 

Wilks' Theorem\cite{Wilks} (see Section \ref{aside}) states that, provided
certain conditions are satisfied,  -2*LLR has a $\chi^2$ distribution;
this appears to contradict our previous conclusion about the distributions of the logarithm of
a likelihood or likelihood ratio
being Gaussian. Do we have a paradox?\footnote{Given that the $\chi^2$ distribution 
is asymptotically Gaussian, it might be thought that the different expectations of Gaussian or 
$\chi^2$ for the LLR are compatible assymptotically. This is incorrect: in the statement about 
$\chi^2$ becoming Gaussian, `asymptotically' refers to the number of degrees of freedom
for the $\chi^2$ being large, and does not refer to the number of observations.} 

\vspace{0.2in}
Section 2 contains some background information on different types of hypotheses and on various likelihood ratios. 
Likelihoods and likelihood ratios for the easily understood exponential distribution are examined
 in Section \ref{Examples}, where the paradox is resolved. 
%The universality of the likelihood as a function of $\tau/\tau_{best}$ is also discussed. 
The relationship between the locations and the widths of the peaks of the pdf's of the LLR is also discussed
in Section \ref{Examples}. A similar analysis for a Gaussian distribution is presented in Section \ref{Gaussian}.
Finally the issue of whether unbinned likelihood provides information on
Goodness of Fit is recalled in Section \ref{GofF}.

\section{Aside on Hypotheses and Likelihood Ratios}
\label{aside}
Statisticians divide hypotheses into those in which there are no free parameters,
and those where there are; they are called Simple and Composite respectively. An 
example of the former would be that the spin of the Higgs boson is 
zero\footnote{In practice, hypotheses in Particle Physics are rarely simple. Even 
if there are no extra physics parameters involved, there are almost always systematic 
nuisance parameters of an experimental nature (e.g. energy calibrations, 
electron identification efficiency, etc.) However, if these systematic effects have only a small 
effect on the analysis, it may be an adequate approximation to treat the hypothesis as 
simple.}, while 
the possible existence of a heavy neutrino with any mass would be composite.
For comparing two simple hypotheses, the Neyman-Pearson Lemma\cite{NP} says
that the likelihood ratio (LR) provides the best way of distinguishing the hypotheses. 

Even when the hypotheses are composite, the LR may provide a reasonable
way of assesing how two different hypotheses compare in describing the data. In 
that case, however, theorems about the distribution of a test statistic for  
simple hypotheses may not apply, and distributions then have to be determined by simulation. 

 A special case of a composite hypothesis occurs when Wilks' Theorem applies (see 
Section \ref{Introd} above). It says that -2*LLR for two hypotheses has a $\chi^2$ 
distribution provided:
\begin{itemize}
\item{The null hypothesis is true.}
\item{The hypotheses are nested. i.e. By a special choice of parameter values for the larger 
hypothesis $H1$, it
can be reduced to the smaller hypothesis $H0$. This automatically requires $H1$ to be composite.}
\item{The extra parameters in $H1$ that are required to reduce it to $H0$ are all uniquely defined,
and are not at the extreme end of their allowed ranges.}
\item{There are enough data for asymptotic approximations to apply.} 
\end{itemize} 
The number of degrees of freedom of the $\chi^2$ distribution is equal to the number of extra 
free parameters in $H1$.

There are several different sorts of LRs that are used in Particle Physics,
and so it is important to specify which we are discussing at any time. Here we consider 
${\mathcal L}(x|H0)/{\mathcal L}(x|H1)$ where the hypothesis $H0$  may or may
not involve free parameter(s), and $H1$ does.
Variants of this LR are discussed in \cite{LW}.

The expected asymptotic distributions of various LLRs are discussed in \cite{CCGV}.

\section{Examples with an exponential distribution}
\label{Examples}
We resolve the paradox with a straightforward example where the pdf is an exponential distribution
\begin{equation}
\label{exp}
p(t|\tau) = (1/\tau) * \exp(-t/\tau)
\end{equation} 
with parameter $\tau$ and observation $t$. This could be for the decay time $t$ of 
a radioactive system with mean life parameter $\tau$ \footnote{In realistic situations, decay time 
distributions are more complicated than this.
Thus background, time resolution, acceptance cuts, etc. would need to be taken into account.
We deal with the idealised case as it is amenable to analytic solution, and hence is useful
for obtaining insights.}. 
Then, with a set of $t_i$ 
observed decay times, the likelihood is
\begin{equation}
{\mathcal L}(\tau) = \Pi\ [(1/\tau) * \exp(-t_i/\tau)]
\end{equation} 
where the product is over the $N$ observations. Then its logarithm is
\begin{equation}
\label{L}
\begin{split}
\ln{\mathcal L}(\tau) &= \Sigma\ [-\ln\tau -t_i/\tau] \\
                   &= -N*\ln\tau - N*\bar t/\tau
\end{split}
\end{equation} 
where ${\bar t}$ is the mean of the observed decay times $t_i$.

\subsection{One simple hypothesis: $\tau = \tau_1$}
First we consider the simple hypothesis, where $\tau$ has a fixed value $\tau_1$.
By the CLT, $\bar t$ is asymptotically Gaussian. Since the mean and variance of the exponential
distribution of eqn. \ref{exp} are $\tau$ and $\tau^2$ respectively, the mean of the 
Gaussian for ${\bar t}$ is $\tau$ and its variance is $\tau^2/N$. 

For fixed $\tau_1$,  $\ln {\mathcal L}$ of eqn. \ref{L} is a linear function of 
 the single variable $\bar t$,  and so it too will be Gaussian\footnote{It is important 
not to be confused between (a) the likelihood as a function of its parameter for a single data set, which 
often is asymptotically 
Gaussian; and (b) the distribution of  the logarithm of the likelihood for a repeated set of experiments, 
which is also Gaussian if the CLT applies. The former involves plotting ${\mathcal L}(\mu)$ against  
$\mu$, while the second is a histogram of the LLR.}.

\subsection{One composite hypothesis: variable $\tau$}
Now we consider a composite hypothesis, with $\tau$ varied to find $\tau_{best}$ that 
maximises $\ln {\mathcal L}$. This yields\footnote{Equation \ref{Trivial}
looks trivially true as it can casually be read as ``The average lifetime is the average lifetime". However 
it does have real content, as the more accurate description is ``Our best estimate of the parameter $\tau$  of the 
exponential function fitted to the data is given by the mean of the $N$ observed decay times." }
\begin{equation}
\label{Trivial}
\tau_{best} = {\bar t}
\end{equation}
Inserting this in eqn. \ref{L} gives 
\begin{equation}
\label{CH}
\begin{split}
\ln{\mathcal L}(\tau_{best}) &= \Sigma\ [-\ln\tau_{best} -t_i/\tau_{best}] \\
                   &= -N*\ln\bar t - N
\end{split}
\end{equation} 
Asymptotically $\bar t$ is still Gaussian but $\ln {\mathcal L}$ involves $\ln \bar t$. However, the distribution of a variable 
$\ln z$ is related to the distribution in $z$ simply by a factor $z$. Since for large $N$, the expected width
of the $\bar t$ distribution becomes narrow, the extra factor of $\bar t$ will not vary too much over the main 
part of the $\bar t$ distribution, and so we expect that asymptotically $\ln \bar t$ will also be 
Gaussian\footnote{It might be thought that rather larger values of $N$ are required for this approximation 
to be good, than for $\bar t$ itself; this is plausible but incorrect. It turns out that for our exponential 
example the Gaussian approximation is better for $\ln\bar t$ than for $\bar t$.}.  

Thus for variable $\tau$,  $\ln {\mathcal L(\tau_{best}})$ is a linear function of 
the single variable $\ln\bar t$,  and so it too asymptotically will be Gaussian.

\subsection{Two simple hypotheses: $\tau = \tau_1$ and $\tau_2$}
\label{TwoHypotheses}
Similarly for the logarithm of the ratio of the likelihoods for two different assumed lifetimes
$\tau_1$ and $\tau_2$ is
\begin{equation}
\ln({\mathcal L}(\tau_1)/{\mathcal  L}(\tau_2))= -N*\ln(\tau_1/\tau_2) - N*{\bar t}*(1/\tau_1 - 1/\tau_2)
\end{equation} 
The LLR for these two simple hypotheses 
is also linearly dependent on ${\bar t}$, and hence the distribution of 
$\ln({\mathcal L}(\tau_1)/{\mathcal L}(\tau_2))$ is again asymptotically Gaussian. We expect 
${\bar t} = \tau_1(1+g/\sqrt N)$ if $\tau = \tau_1$, or $\tau_2(1 + h/\sqrt N)$ if $\tau  = \tau_2$,
where $g$ and $h$ are standard Gaussians of mean zero and unit variance.

Writing $\tau_1 = (1+k)\tau_2$, where $k$ is small if $\tau_1 \sim \tau_2$, we obtain for 
the expected values of the likelihood ratio, assuming that $\tau_1$ is the true value
\begin{equation} 
\label{LLR1}
\begin{split}
[\ln({\mathcal L} (\tau_1)/{\mathcal L} (\tau_2))]_{\tau_1} &=
-N\ln (1+k) - N(1 + g/\sqrt N)(1-(1+k))\\
&\sim -N[(k - k^2/2  +  k^3/3 + ...) - k] + Nkg/\sqrt N\\
&\sim [Nk^2(1/2 -  k/3 + ....)] + [\sqrt N kg]
\end{split}
\end{equation}
Alternatively, if $\tau_2$ is the true value
\begin{equation}
\label{LLR2} 
\begin{split}
[\ln(\mathcal L (\tau_1)/\mathcal L (\tau_2)]_{\tau_2} &=
-N \ln (1+k) -N(1 + h/\sqrt N)(1/(1+k)-1)\\
&\sim -N[((k-k^2/2 + k^3/3  ...) + (1-k + k^2  -k^3 ....) -1] + N(k - k^2 ...)h/\sqrt N\\
&\sim [-Nk^2(1/2 - 2k/3 +....)] + [\sqrt Nhk(1 - k + ...)]
\end{split}
\end{equation}
Because $g$ and $h$ are standard Gaussians, the first term on the right hand side of eqns \ref{LLR1} 
and \ref{LLR2} gives the location of the peak, and the second term determines its width. 
Thus for the two similar pdfs (exponentials with $\tau_1 \sim \tau_2$), $k$ is small and the distributions of the 
LLR for the two hypotheses peak at symmetric values $\pm T_E$ of the LLR, equidistant from zero, and their widths $w_E$
are equal. Furthermore with close values of $\tau$, the widths and locations of
the LLR distributions for the two hypotheses are related by\footnote{Sometimes the distributions of
-2*LLR are considered. In that case, the locations $\pm T'$ and widths $w'$ are related by $w' = 2\sqrt T'$.} 
\begin{equation}
\label{wT}
w_E = \sqrt{2T_E} 
\end{equation}
%this is not true if the values of $\tau_1$ and $\tau_2$ are not close.

This result is reminiscent of the situation of trying to distinguish the two hypotheses of normal and 
inverted hierarchies for the masses of neutrinos of different flavours\footnote{In the neutrino
mass hierarchy situation, the data pdf's for the two hypotheses have different functional forms.
This is not so in our simple example where both $H0$ and $H1$ correspond to exponential 
distributions (with different values of $\tau$).}, which result in small differences 
of observed neutrino spectra. The claim there is that the LLR distributions for these two hypotheses 
tend to be Gaussians with locations $\pm T$ and widths given by eqn \ref{wT} \cite{Tan,Blennow}.

When the pdf's are more different, the higher order terms of the expansions of eqns. 
\ref{LLR1} and \ref{LLR2} are relevant, and the symmetry of locations and equality of widths no longer holds. 
(Compare, for example, the case of trying to distinguish spin-parity $0^+$ from $1^+$
for the Higgs boson - see fig. 5 of ref. \cite{CMS_spin}).

Another interesting result is that, for small $k$ (i.e. $\tau_1 \sim \tau_2$),
\begin{equation}
\label{widths}
2T_E/w_E \sim \sqrt N k =\sqrt N (\tau_2 - \tau_1)/\tau_1
\end{equation}
That is, the separation of the peaks of the distributions  of the LLR for the two hypotheses is approximately
equal to the difference in the mean lifetimes, divided by the uncertainty $\tau/\sqrt N$ in determining a lifetime 
from $N$ events.

\subsection{Nested hypotheses and Wilks' Theorem for Exponentials}
Wilks' Theorem applies to the situation where $H0$ (e.g. $\tau = \tau_{gen}$) is true, and $H1$ is just an 
extended version of $H0$ with extra parameter(s) - see Section \ref{aside}; 
in our exponential example, $\tau$ is allowed to vary. Then $H1$'s 
likelihood must be at least as large (as good) as that for $H0$, but if $H0$ is true, the freedom due to the 
extra parameter(s) of $H1$ is not required, and we expect $-2*\ln({\mathcal L}_0/{\mathcal L}_1)$
to be small. Wilks' Theorem gives us a way of quantifying this. In contrast, if $H0$ is not true, -2*LLR can be large.
\vspace{0.1in}

Here we look at $\Delta\ln\mathcal L = \ln\mathcal L (\tau_{test}) - \ln\mathcal L  (\tau_{best})$, where the data consist 
of decay times generated according to an exponential distribution with $\tau = \tau_{gen}$, and the two likelihoods
are respectively for specific tested values and for its best value $\tau_{best}$ for that data set. 

If we set $\tau_{test} = \tau_{gen}$, we expect $-2\Delta\ln\mathcal L$ to have a small positive (or zero) value.
In fact, asymptotically the conditions for Wilks' Theorem are satisfied, in which case its expected distribution is 
$\chi^2$ with one degree of freedom.

We can verify this because from equations \ref{L} and \ref{CH} we obtain
\begin{equation}
\begin{split}
\Delta\ln{\mathcal L} &= (-N\ln\tau_{test} - N{\bar t}/\tau_{test}) -
(-N\ln{\bar t} -N) \\
&=N(\ln{\bar t}/\tau_{test} -({\bar t}/\tau_{test} -1)) 
\end{split}
\end{equation}
We expect ${\bar t} =  \tau_{best}$ to be Gaussian distributed about $\tau_{gen}$ with width $\tau_{gen}/{\sqrt N}$ 
(see first paragraph of section \ref{TwoHypotheses}) i.e.
\begin{equation}
\tau_{best} = \tau_{gen}(1 + g/{\sqrt N})
\end{equation}
where $g$ is a standard Gaussian random variable, and so
\begin{equation}
\Delta\ln\mathcal L  = N(\ln(1+g/\sqrt N) - g/\sqrt N)
\end{equation}
For large N we can expand $\Delta\ln\mathcal L$ in powers of $g/\sqrt N$, to obtain
\begin{equation}
\Delta\ln\mathcal L = N(-g^2/2N + .....)
\end{equation}
The leading term in $g/{\sqrt N}$ vanishes, and so asymptotically $-2\Delta\ln\mathcal L = g^2$, 
the square of a standard Gaussian variable, and hence its expected distribution is indeed
$\chi^2$ with one degree of freedom. It is the cancellation of the leading term in $g/\sqrt N$
which results in the LLR not having a Gaussian distribution in this case.

\vspace{0.2in}

Not specifically related to the `Gaussian or $\chi^2$' paradox, but interesting in its own right is
the fact that  when we obtain the likelihood function for the exponential's parameter $\tau$ from $N$ observed decay times, 
the difference $\Delta \ln{\mathcal L}$ between the likelihood and its best value is a universal 
function of $\tau/\tau_{best} = \lambda$. From equation \ref{CH} we obtain
\begin{equation}
\begin{split}
\Delta \ln{\mathcal L} &=\ln{\mathcal L}(\tau) - \ln{\mathcal L}(\tau_{best})\\
%&= (-N*\ln\tau - N*\bar t/\tau) - (-N*\ln\bar t - N) \\
&= N(\ln {\bar t}/\tau +1 - {\bar t}/\tau)\\
&= N(1 - 1/\lambda - \ln\lambda)
\end{split}
\end{equation}  
thus demonstrating the assertion. This implies that the uncertainty on $\tau/\tau_{best}$ is 
independent of $\tau_{best}$ and of the individual data values (apart from their average
$\bar t$), and depends only on $N$. For example,
if the uncertainty on $\tau$ is estimated as 
$(-d^2\ln\mathcal L/d\tau^2)^{-1/2}$, we obtain $\sigma_{\tau} = \tau/\sqrt N$.

%%%%%%%%%%%%%%%%%%%%%%%%%%%%%%%%%%%%%%%%%%%%%%%%%%%%%%%%%%%%%%%%%%
%%%%%%%%%%%%%%%%%%%%%%%%%%%%%%%%%%%%%%%%%%%%%%%%%%%%%%%%%%%%%%%%%%

% Now for Gaussian pdf

\section{Examples with Gaussian distribution}
\label{Gaussian}
Here  we follow the same situations as for Section \ref{Examples}, but this time using a Gaussian pdf
\begin {equation}
\label{normal}
p(x|\mu)  =  \frac{1}{\sqrt{2\pi}\sigma}exp(-0.5*(x-\mu)^2/\sigma^2)
\end{equation}
i.e. The pdf  for the data $x$ is a Gaussian centred at $\mu$ and with fixed 
width $\sigma$. Then for a set of $N$ observations $x_i$, the log-likelihood is 
\begin{equation}
\begin{split}
\ln{\mathcal L}(\mu)  &= -\Sigma(0.5(x_i - \mu)^2/\sigma^2) + C  \\
&=-0.5  N({\bar x} - \mu)^2/\sigma^2 + C
\end{split}
\end{equation}
where the constant $C = -N\ln(\sqrt{2\pi}\sigma)$.
Thus, as in the exponential case,  the log-likelihood depends on the data only though its mean
$\bar x = \Sigma x_i/ N$. 

\subsection{One simple hypothesis: $\mu = \mu_1$}
For fixed $\mu_1$, the log-likelihood is thus given by
\begin{equation}
\label{G1S}
\ln{\mathcal L}(\mu_1) = -\Sigma(0.5(x_i - \mu_1)^2/\sigma^2) + C
\end{equation}
Since with $x$ having a Gaussian distribution, that for $(x - \mu)^2$ is specified, and so
by the CLT, $\ln{\mathcal L}$ will have a Gaussian distribution asymptotically.

An alternative derivation uses the fact that each $x_i$ is independently Gaussian distributed 
with mean $\mu_1$ and variance $\sigma^2$. Thus each $(x_i - \mu_1)/\sigma$ is
distributed as a standard Gaussian, with zero mean and unit variance. Then,
apart from the constant $C$, $-2\ln{\mathcal L}(\mu_1)$ is distributed as $\chi^2$ with 
$N$ degrees of freedom. Since asymptotically (i.e. large $N$), $\chi^2_N$ tends to a Gaussian
distribution, this also applies to the log-likelihood.

\subsection{One composite hypothesis: variable $\mu$}
We now consider a composite hypothesis, with $\mu$ varied to find $\mu_{best}$ that maximises 
$\ln{\mathcal L}$. As with the exponential case, the best value of the parameter is equal to
the mean of the data, i.e. $\mu_{best}  =  \bar x $. 
Inserting this into eqn. \ref{G1S} yields
\begin{equation}
\begin{split}
\ln{\mathcal L}(\mu_{best}) 
%&= -0.5\Sigma(x_i -\mu_{best})^2/\sigma^2 + C \\
  &= - 0.5 \Sigma (x_i - \bar x)^2/\sigma^2 + C
\end{split}
\end{equation}

\subsection{Two simple hypotheses: $\mu = \mu_1$ and $\mu_2$}
Similarly  the LLRs for two different assumed central values
$\mu_1$ and $\mu_2$ is
\begin{equation}
\label{TwoSimple}
\begin{split}
\ln({\mathcal L}(\mu_1)/{\mathcal  L}(\mu_2)) 
&= -\frac{1}{2\sigma^2}[-2\Sigma x_i \mu_1 +2\Sigma x_i\mu_2 +\mu_1^2 -\mu_2^2]  \\
&=-\frac{N (\mu_2 - \mu_1)}{\sigma^2} [\bar x - (\mu_1  + \mu_2)/2]  
% -N*\ln(\tau_1/\tau_2) - N*{\bar t}*(1/\tau_1 - 1/\tau_2)
\end{split}
\end{equation} 
where the terms quadratic in $x_i$ and the constant $C$ have cancelled between the two log-likelihoods.
The LLR for these two simple hypotheses 
is linearly dependent on ${\bar x}$, and hence the distribution of 
$\ln({\mathcal L}(\mu_1)/{\mathcal L}(\mu_2))$ is again Gaussian. But here there is no need to invoke the 
CLT, since the sum of any number of random variables drawn from identical Gaussians (i.e. equal widths and 
equal central values)  has a Gaussian distribution. We expect 
%${\bar t} = \tau_1(1+g/\sqrt N)$ if $\tau = \tau_1$, or $\tau_2(1 + h/\sqrt N)$ if $\tau  = \tau_2$,
${\bar x} = \mu_1 +g\sigma/\sqrt N$ if $\mu = \mu_1$, or $\mu_2 + h\sigma/\sqrt N$ for $\mu_2$,
where $g$ and $h$ are standard Gaussians of mean zero and unit variance.

%Writing $\tau_1 = (1+k)\tau_2$, where $k$ is small if $\tau_1 \sim \tau_2$, we obtain for 
%the expected values of the likelihood ratio, 
Assuming that $\mu_1$ is the true value
\begin{equation} 
\label{LLR3}
\begin{split}
[\ln({\mathcal L} (\mu_1)/{\mathcal L} (\mu_2))]_{\mu_1} 
%-N\ln (1+k) - N(1 + g/\sqrt N)(1-(1+k))\\
%&\sim -N[(k - k^2/2  +  k^3/3 + ...) - k] + Nkg/\sqrt N\\
%&\sim [Nk^2(1/2 -  k/3 + ....)] + [\sqrt N kg]
&=-\frac{N(\mu_2 - \mu_1)}{\sigma^2}[(\mu_1 + g\sigma/\sqrt N) - (\mu_1 +\mu_2)/2]   \\
&= \frac{N}{\sigma^2} (\mu_2 - \mu_1)^2/2 - \frac{N}{\sigma^2}(\mu_2-\mu_1)\frac{g\sigma}{\sqrt N} 
\end{split}
\end{equation}
Because $g$ is a standard Gaussian, the first term above gives the location $T_G$ of the peak
of the LLR distribution, while the coefficient
of $g$ in the second term gives its width $w_G$ i.e.
\begin{equation} 
\label{GTw}
T_G = \frac{N}{\sigma^2} (\mu_2 - \mu_1)^2/2;\ \ \ \ \ w_G = \sqrt N |\mu_2 - \mu_1|/\sigma%;  and \ w_G = \sqrt{2T_G}
\end{equation}  
Alternatively, if $\mu_2$ is the true value.
the peak is at  $-T_G$, and the width is as in eqn \ref{GTw}.
%while for $\mu_2$
%\begin{equation}
%\label{LLR4} 
%\begin{split}
%[\ln(\mathcal L (\mu_1)/\mathcal L (\mu_2)]_{\mu_2} &=
% -\frac{N}{\sigma^2} (\mu_2 - \mu_1)^2/2  = -T_G,
%-N \ln (1+k) -N(1 + h/\sqrt N)(1/(1+k)-1)\\
%&\sim -N[((k-k^2/2 + k^3/3  ...) + (1-k + k^2  -k^3 ....) -1] + N(k - k^2 ...)h/\sqrt N\\
%&\sim [-Nk^2(1/2 - 2k/3 +....)] + [\sqrt Nhk(1 - k + ...)]
%\end{split}
%\end{equation}

Thus for the two Gaussian pdfs  the distributions of the 
LLR for the two hypotheses peak at symmetric values $\pm T_G$, equidistant from zero, and their widths $w_G$
are equal. Furthermore, the widths and locations of
the LLR distributions for the two hypotheses are related by 
\begin{equation}
\label{GTwrelation}
w_G = \sqrt{2T_G} 
\end{equation}
%this is not true if the values of $\tau_1$ and $\tau_2$ are not close.
as in the case of two exponentials (but there the values of $\tau_1$ and $\tau_2$ need to be close
and we require asymptotic data for the relationships concerning the distributions' locations and widths to apply).
Equations \ref{GTw} and \ref{GTwrelation} are true for any values of the parameters $\mu_1$ and $\mu_2$.
and do not require asymptotic data. 

It is also worth noting that 
\begin{equation}
2T_G/w_G = |\mu_2 - \mu_1|\sqrt N/\sigma
\end{equation}
i.e the separation of the peaks of the LLR distributions, divided by their width, is equal to the 
separation of the original Gaussian pdfs, divided by the uncertainty $\sigma/\sqrt N$ in their experimental location
(cf. eqn. \ref{widths}).

\subsection{Nested hypotheses and Wilks' Theorem for Gaussians}
%Wilks' Theorem applies to the situas in the case ation where $H0$ (e.g. $\tau = \tau_{gen}$) is true, and $H1$ is just an 
%extended version of $H0$ with extra parameter(s) - see Section \ref{aside}; 
%in our exponential example, $\tau$ is allowed to vary. Then $H1$'s 
%likelihood must be at least as large (as good) as that for $H0$, but if $H0$ is true, the freedom due to the 
%extra parameter(s) of $H1$ is not required, and we expect $-2*\ln({\mathcal L}_0/{\mathcal L}_1)$
%to be small. Wilks' Theorem gives us a way of quantifying this. In contrast, if $H0$ is not true, -2*LLR can be large.
%\vspace{0.1in}

Here we look at $\Delta\ln\mathcal L = \ln\mathcal L (\mu_{test}) - \ln\mathcal L  (\mu_{best})$, where the data $x$
are generated according to a Gaussian distribution with $\mu = \mu_{gen}$, and the two likelihoods
are respectively for the specific tested value  and for its best value $\mu_{best}$ for that data set. 

If we set $\mu_{test} = \mu_{gen}$, we expect $-2\Delta\ln\mathcal L$ to have a small positive (or zero) value.
The conditions for Wilks' Theorem are satisfied, in which case its expected distribution is 
$\chi^2$ with one degree of freedom.

We can verify this because from equation \ref{TwoSimple}
\begin{equation}
\begin{split}
%\Delta\ln{\mathcal L} &= (-N\ln\mu_{test} - N{\bar x}/\mu_{test}) -
%(-N\ln{\bar t} -N) \\
%&=N(\ln{\bar t}/\tau_{test} -({\bar t}/\tau_{test} -1)) 
\Delta\ln{\mathcal L} &= \ln{\mathcal L}(\mu_{gen}) - \ln{\mathcal L}(\mu_{best}) \\
&= -\frac{N}{\sigma^2}(\mu_{best} - \mu_{gen})({\bar x} - (\mu_{best} + \mu_{gen})/2)
\end{split}
\end{equation}
We expect ${\bar x} =  \mu_{best}$ to be Gaussian distributed about $\mu_{gen}$ with width $\sigma/{\sqrt N}$ 
%(see first paragraph of section \ref{TwoHypotheses}) 
i.e.
\begin{equation}
\mu_{best} = \mu_{gen} +l\sigma/\sqrt N
\end{equation}
where $l$ is a standard Gausian random variable, and so
\begin{equation}
\begin{split}
\Delta\ln\mathcal L  &= -\frac{N}{\sigma^2}\frac{l\sigma}{\sqrt N}[0.5(\mu_{gen} +l\sigma/\sqrt N) - 0.5\mu_{gen}] \\
&=-l^2/2
\end{split}
\end{equation}
%For large N we can expand $\Delta\ln\mathcal L$ in powers of $g/\sqrt N$, to obtain
%\begin{equation}
%\Delta\ln\mathcal L = N(-g^2/2N + .....)
%\end{equation}
%The leading term in $g/{\sqrt N}$ vanishes, and so asymptotically 
Thus $-2\Delta\ln\mathcal L = l^2$, 
the square of a standard Gaussian variable, and hence its expected distribution is indeed
$\chi^2$ with one degree of freedom. Again this result involves neither approximation
nor the need for asymptotic data.

\section{Unbinned likelihoods and Goodness of Fit}
\label{GofF}
While considering likelihood functions for exponentials and for Gaussians, we discuss an interesting property.

The likelihood method asserts that the best value of a parameter is obtained by maximising the likelihood 
with respect to the parameter. It thus might be thought that large likelihood values are better than smaller
ones, and hence that large likelihood values are indicative of a better goodness of fit between data 
and the chosen parametric form. 
 
For unbinned likelihoods, 
this is incorrect. The likelihood is a measure of the probability density for obtaining  the given
data set, which is {\bf fixed}; it uses these probabilities as the parameter is varied. In contrast, Goodness of 
Fit involves the probabilities of ${\bf different}$ data sets, for a fixed value of the parameter.

The inability of the unbinned maximum likelihood to distinguish between data which does have the expected
distribution and data which does not is illustrated for the exponential example by eqn \ref{CH}. This shows 
that ${\mathcal L}(\tau_{best})$ depends on the data only through the average of their decay times ${\bar t}$. Thus
any sets of $N$ observations  which happen to have the same value of $\bar t$ will have the identical value of
the maximum likelihoods; these data  sets could be distributed as expected for an exponential distribution, or
one where all the decays occurred at the identical time. Thus the value of ${\mathcal L}(\tau_{best})$ is
incapable of distinguishing between an acceptable data set, and one which is very strongly in disagreement 
with the exponential decay hypothesis (see also ref. \cite{Heinrich}).

The Gaussian case provides an even stronger example. We want to test goodness of fit of the Gaussian 
distribution of eqn. \ref{normal} with fixed width with two data sets: the first has the individual unbinned 
$x_i$ distributed as expected for eqn. \ref{normal}, while the second has all the data $x_i$ equal to  $\mu$. This 
second set is not compatible with our chosen fixed width Gaussian, but clearly results in a larger value for the 
unbinned likelihood than the first data set. The data which is less compatible with the hypothesis gives 
a larger likelihood. The value of the unbinned likelihood is clearly not a good measure of Goodness of Fit. 

\vspace{0.1in}

Baker and Cousins\cite{BC} provide a prescription for obtaining Goodness of Fit information 
for a likelihood approach to a histogram, i.e. for a binned likelihood approach\footnote{In
contrast, the likelihood-ratio Wilks' Theorem can use unbinned data.}. But even here, 
the likelihood alone cannot be used. A reason for this is that histogram bins with $n_{obs}$ observed and $\lambda_{pred}$
predicted events being 1 and 1.0 repectively, or with $n_{obs} = 100$ and $\lambda_{pred} = 100.0$, both have 
perfect agreement for data and prediction. However, because the Poisson distribution 
\begin{equation}
p(n|\lambda) = e^{-\lambda} \lambda^n/n!
\end{equation}
is much wider for $\lambda =100.0$ than for $\lambda = 1.0$, the likelihoods are very different (0.37 for 
$\lambda = 1.0$, but only 0.04 for $\lambda = 100.0$). Baker and Cousins overcome the problem by 
considering instead the likelihood ratio ${\mathcal L}(\lambda|n)/{\mathcal L}(n|n)$, where 
${\mathcal L}(n|n)$ is the likelihood for the `saturated' model i.e. where $\lambda$ is chosen as the value
which maximises ${\mathcal L}(\lambda|n)$ for the observed $n$. See \cite{BC} for further details.

\section{Conclusion} 
The Table summarises the various results for the exponential and for the Gaussian pdf's. 

%{\hskip -3.0cm}
\begin{table}
\caption{Summary of examples with exponential or Gaussian pdfs.}
%\begin{adjustbox}{center}
\begin{tabular} {|c|c|c|}
\hline
        & Exponential & Gaussian \\ \hline
pdf &  $p(t|\tau) = (1/\tau) * \exp(-t/\tau)$  &  
$p(x|\mu)  =  \frac{1}{\sqrt{2\pi}\sigma}exp(-0.5*(x-\mu)^2/\sigma^2)$	\\  \hline
Best value of param   &   $\tau = {\bar t}$   & $\mu = {\bar x}$  \\    \hline
$\ln{\mathcal L}(\tau\ or\ \mu)$ &  $\ln{\mathcal L}(\tau) = -N*\ln\tau - N*\bar t/\tau$   &  
$\ln{\mathcal L}(\mu) =-0.5  N({\bar x} - \mu)^2/\sigma^2 -N\ln(\sqrt{2\pi}\sigma)$       \\  
    & i.e. linear in ${\bar t}$   &  i.e. linear in $({\bar x} - \mu)^2$ \\ \hline
$\ln{\mathcal L}(\tau_{best}\  or\ \mu_{best})$ &  $\ln{\mathcal L}(\tau_{best}) = -N*\ln\bar t - N$   & 
$\ln{\mathcal L}(\mu_{best})  = - 0.5 \Sigma (x_i - \bar x)^2/\sigma^2 -N\ln(\sqrt{2\pi}\sigma)$  \\  
    & i.e linear in $\ln {\bar t}$   &   i.e. linear in $\Sigma(x_i - {\bar x})^2$ \\ \hline
LLR for 2 simple hyps      &  For $\tau_1\sim\tau_2$   &   For any $\mu_1$ and $\mu_2$      \\
                           & peaks at $\pm T_E$, equal widths $w_E$ & peaks at $\pm T_G$, equal widths $w_G$  \\  
											  		& $w_E^2 = 2T_E$   &    $w_G^2 = 2T_G$  \\  
& $2T_E/w_E \sim |\tau_1 - \tau_2|/(\tau_1/\sqrt N)$  &  $2T_G/w_G = |\mu_1 - \mu_2|/(\sigma/\sqrt N)$  \\  \hline
$-2\ln({\mathcal L}(\tau\ or\ \mu)/{\mathcal L}_{best})$    & Asymptotically $\chi^2$    &   Always $\chi^2$    \\
\hline
\end{tabular}
\label{table:conclusions}
%\end{adjustbox}
\end{table}
		
The straightforward example of an exponential distribution for the pdf illustrates that
the logarithm of the likelihood  will asymptotically be Gaussian. 
%provided that there are no free parameters in the relevant hypotheses i.e. that the hypotheses are Simple. 
This is also true for the logarithm of the likelihood ratio for two simple hypotheses each with a fixed value of
$\tau$. If the values of $\tau$ for the two hypotheses are close to each other, the peaks of the Gaussian distributions 
of the LLR for the two hypotheses will be symmetrically located at $\pm T$ and with widths equal to $\sqrt{2T}$.

However, for the LLR in the case of nested hypotheses where Wilks Theorem applies and assuming 
that the smaller hypothesis is true, it is a cancellation between the leading term in each of 
the individual likelihoods which results in -2*LLR being distributed as a $\chi^2$, rather than Gaussian.
There is thus no paradox.

The above results are also true for Gaussian pdf's, but the results are exact and no approximation is necessary.

It is interesting to speculate whether the above results about the positions and widths of
the LLR distributions have a wider applicability than for just exponential and Gaussian pdf's.

\vspace{0.3in}
I wish to thank Bob Cousins, David van Dyk and Sara Alegri for illuminating discusssions.

%\balance
%

%

\begin{thebibliography}{99}
\bibitem{Wilks} S. S. Wilks, `The large-sample distribution of the likelihood ratio for testing 
composite hypotheses', Ann. Math. Statist. {\bf 9} (03, 1938) 60.
\bibitem{NP} J. Neyman and E. S. Pearson, `On the problem of the most efficient tests
of statistical hypotheses', Phil. Trans. Roy. Soc. Series A {\bf 231} (1933) 289.
\bibitem{LW} L. Lyons and N. Wardle, `Statistical issues in searches for new phenomena in
High Energy Physics', to be published (2017).
\bibitem{CCGV} G. Cowan, K. Cranmer, E. Gross and O. Vitells, `Asymptotic formulae for 
likelihood-based tests of new physics', Eur. Phys. J. {\bf C71} (2011) 1554.
\bibitem{Tan} X. Qian et al., `Statistical evaluation of experimental determinations of 
neutrino mass hierarchy, Phys. Rev. {\bf D86} (2012) 113011.
\bibitem{Blennow} M. Blennow et al., `Quantifying the sensitivity of oscillation experiments 
to the neutrino mass ordering', JHEP {\bf 03} (2014) 028.
\bibitem{CMS_spin} CMS Collaboration, V. Khachatryan et al., `Constraints on the spin-parity 
and anomalous HVV couplings of the Higgs boson in proton collisions at 7 and 8 TeV', 
Phys. Rev. {\bf D92} (2015) 101803.
\bibitem{Heinrich} J. Heinrich, 'Can the likelihood function be used to measure 
goodness of fit?', CDF note 5639 (2001)
\bibitem{BC} S. Baker and R. D. Cousins, `Clarification of the use of chi-square and
likelihood functions in fits to histograms', Nucl. Instrum. Meth. ${\bf 221}$ (1984)
437.
\end{thebibliography}
\end{document}